\documentclass[preprint]{revtex4}  
\usepackage{amsfonts}
\usepackage{amsmath}
\usepackage{cancel}
\usepackage{graphicx}
\usepackage{mathcomp}
\usepackage{xcolor}
\usepackage{ulem}
\usepackage{enumitem}

\newif\iffigures
\figurestrue

\renewcommand{\vec}{\boldsymbol}

\renewcommand{\vec}{\boldsymbol}

\def\undertilde#1{\mathord{\vtop{\ialign{##\crcr
$\hfil\displaystyle{#1}\hfil$\crcr\noalign{\kern1.5pt\nointerlineskip}
$\hfil\widetilde{}\hfil$\crcr\noalign{\kern1.5pt}}}}}

\begin{document}
\title{The Multipole Resonance Probe: Simultaneous Determination of Electron Density and Electron Temperature Using Spectral Kinetic Simulation}

\author{Junbo Gong$^{1}$, Michael Friedrichs$^{2}$, Jens Oberrath$^{2}$, and Ralf Peter Brinkmann$^{1}$}
\affiliation{$^{1}$ Ruhr University Bochum\\ Department for Electrical Engineering and Information Technology
\\Institute of Theoretical Electrical Engineering\\ D-44801 Bochum, Germany\\
$^{2}$ South Westphalia University of Applied Science\\
Department of Electrical Power Engineering\\
Modeling and Simulation\\
D-59494 Soest, Germany}

\date{\today}
\renewcommand{\abstractname}{}

\begin{abstract}
\vspace*{0.5cm}
\textcolor[rgb]{0.1,0.1,0.1}{The \textit{Multipole Resonance Probe (MRP)} is a radio-frequency driven probe of a particular \linebreak spherical design, which is suitable for the supervision and control of low-temperature plasmas. The investigation of a spectral kinetic model of the MRP is presented and discussed in this paper. The importance of the kinetic effects was introduced in a previous study of the spectral kinetic model of the idealized MRP. Such effects particularly dominate the energy loss in a low-pressure regime, which are unfortunately absent in the Drude model. With the help of a spectral kinetic scheme, those energy losses can be predicted. It allows obtaining the electron temperature $T_{\rm{e}}$ from the full width at half maximum $\Delta \omega$ in the simulated resonance curve. Simultaneously, the electron density $n_{\rm{e}}$ can be derived from the simulated resonance frequency. Good agreements in the comparison between the simulation and the measurement demonstrate the suitability of the presented model.
}
\end{abstract}

\centerline{ }

\maketitle
\thispagestyle{empty}
\pagebreak

\newpage
\section{Introduction}
 \pagenumbering{arabic}
Supervision and control of plasmas in industrial applications are of notable importance. Nevertheless, diagnostic systems are still challenging due to the numerous industrial \linebreak requirements. One of the industry-compatible candidates is the active plasma resonance spectroscopy (APRS) due to the feature that the plasmas resonate near the electron plasma frequency $\omega_{\rm{pe}}$ \cite{Tonk and Langmuir}. This time-honored concept was initially investigated back in 1960 and it is applied and discussed in many different designs \cite{TakayamaMiyazaki1960,Levitskii1963,Buckley1966,Stenzel1976,PiejakGodyak2004,Dine2005,Scharwitz2009,XuSugai2009,LiLiu2010,WangLiu2011,LinagSugai2011}.
In general, the idea of APRS is that a radio-frequency signal in the GHz range is coupled into the plasma via a probe. In order to determine some important plasma parameters such as electron density $n_{\rm{e}}$, and electron temperature $T_{\rm{e}}$, the resonance is recorded and then analyzed in a mathematical model.

\begin{figure}[!htb]	
 \includegraphics[width=0.7\textwidth]{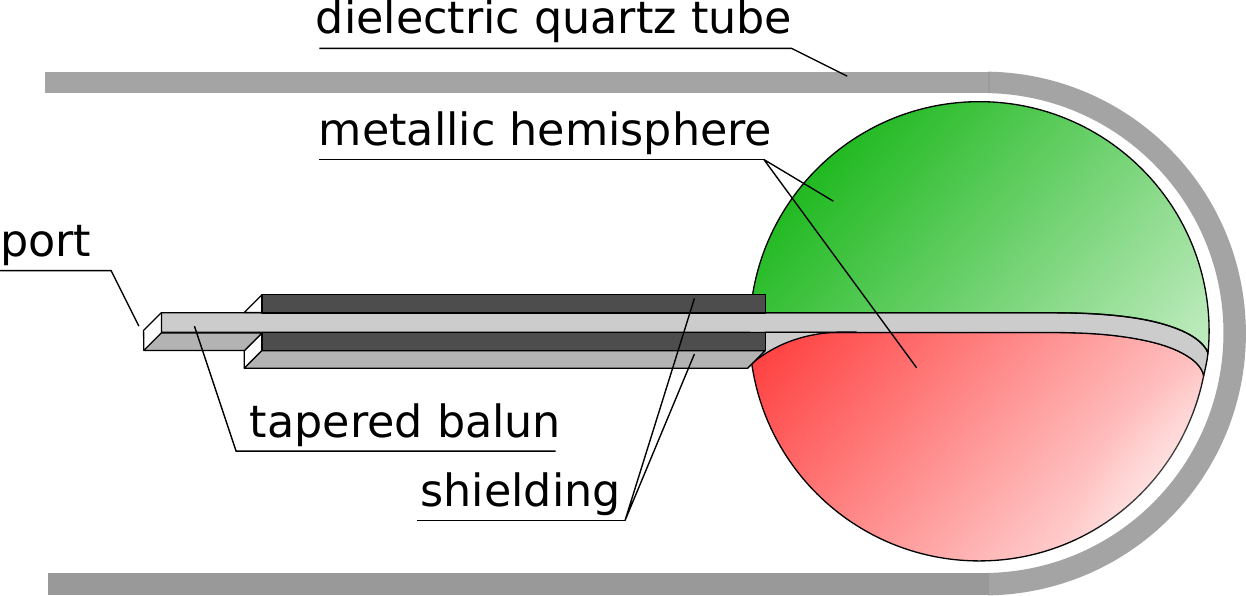} 
  \caption{Sketch of the prototype of the MRP: The probe consists of two metallic hemispheres with a total diameter $8\,\rm{mm}$. It is symmetrically driven via a tapered balun transformer and can be covered in a cylindrical quartz tube.}
 \label{fig:MRP_SCHEMATIC}
\end{figure}

Based on APRS, the Multipole Resonance Probe (MRP) is proposed as a process-compatible diagnostic system~\cite{Lapke2008}. As shown in Fig.~\ref{fig:MRP_SCHEMATIC}, the setup of MRP is electrically antisymmetric and geometrically symmetric: It consists of two metallic hemispheres and a holder. The whole system is immersed in the plasma and fed with a radio-frequency signal approximately between $100\,\rm{MHz}$ and $10\,\rm{GHz}$. The resulting response of the plasma is then recorded and evaluated.

Over the last decade, many investigations of the MRP have been conducted to understand the resonance behavior of the plasma, including analytical and numerical studies based on the Drude model \cite{Lapke2011,SchulzRolfes2014,SchulzStyrnollOberrath2013,Oberrath2021sum}. However, it has been revealed that the validation of the results in the Drude model is limited due to the absence of the kinetic effects. Specifically, the energy transport cannot be described since the only energy loss in the Drude model is due to the collisions. Nevertheless, these effects are influential in the low-pressure regime, which is emphasized in \cite{Lapke2013,Jens_APRS,Jens_PEP,Jens_IP,Jens2020} via functional analytic (Hilbert space) methods. In a recent paper \cite{Gong}, the pure kinetic effects are demonstrated using the spectral kinetic model. The energy loss in the collisionless case can be captured, which is caused by the escape of the free particles in the influenced domain. 

In fact, the electron density $n_{\rm{e}}$, the electron temperature $T_{\rm{e}}$, and the collision frequency are essential for characterizing plasmas. Therefore, the concept for simultaneous determination of electron density and electron temperature was proposed back in 2008~\cite{Lapke2008}. The electron density can be evaluated from the resonance frequency $\omega_{\rm{res}}$, and the relation between the electron temperature to the energy loss is discussed, but the formula is unclarified~\cite{SchulzStyrnollLapke2012}.
With the spectral kinetic model, it is possible to define the formula for the determination of the electron temperature from the full width at half maximum (FWHM)~$\Delta\omega$ of the resonance curve. Hence, only one measurement is needed to obtain these plasma parameters.

Apparently, the spectral response of the linear system is the essence of this work. Notably, the calculations in the frequency domain are computationally intensive with regards to the convergence of a sequence of periodic functions. Therefore, an efficient approach is required. In the spectral kinetic scheme, an impulse signal is provided as the input of the system, then the charge on the electrodes in the time domain are recorded as the output. Then the impulse response can be converted in the frequency domain by applying a Fourier transform.

The outline of the manuscript is as follows. In Section 2, a simplified version of the MRP is introduced, and the spectral kinetic model is theoretically investigated. In Section 3, the simulation results of the spectral kinetic model are presented in detail. The cases with a variation of the electron plasma frequency, in addition to the different electron temperatures, are discussed. Particularly, the relation between the electron temperature and the collisionless kinetic effects is emphasized. Then the collisions between the electrons and the neutral atoms are included to have a complete kinetic investigation. Eventually, good agreements in the comparison between the simulation and the measurement are demonstrated. Section 4 concludes the paper with a brief discussion of these results.

\newpage
\section{Spectral kinetic model of the Ideal MRP}
\begin{figure}[!htb]
 \includegraphics[width=0.55 \textwidth]{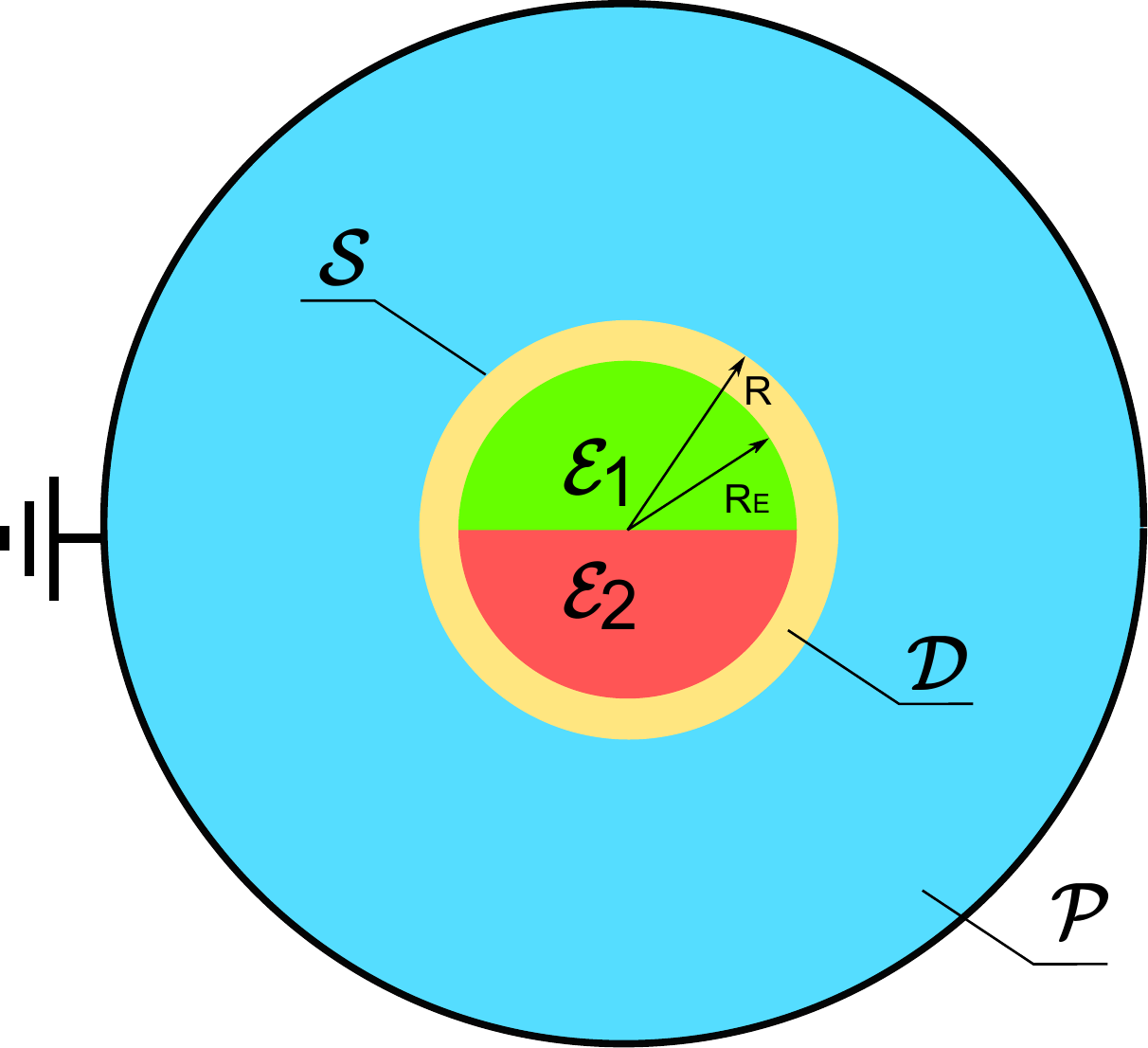} 
  \caption{Illustration of the ideal axially symmetrical model of the MRP covered by the dielectric $\mathcal{D}$ inside a plasma $\mathcal{P}$: the radius of the electrodes $\mathcal{E}_{1,2}$ is $R_{\rm{E}}$ and the radius of the probe is $R$.}
 \label{fig:IMRP}
\end{figure}

Considering the complexity of particle-based simulations, an ideal assembly of the MRP is proposed to enhance the efficiency, which is a highly optimized version of the MRP. The holder can be neglected in the ideal MRP (IMRP) to provide a spherical symmetry in the model. As shown in Fig.~\ref{fig:IMRP}, the IMRP consists of two antisymmetrically powered hemispheres covered with a dielectric. In~\cite{C.SchulzIEEE2014}, the simulation results of an ideal and non-ideal model are very similar, solely a small shift of the resonance frequency appears. It indicates that the influence of the holder can be neglected. Hence, regarding the mathematical description of the system, the spherical form of the IMRP is advantageous for the theoretical investigation. Thus, due to the limitation of the Drude model, the spectral kinetic simulation of the IMRP is presented to explain the influence of the kinetic effects on the resonance structure.

In the spectral kinetic scheme, the plasma around the probe is treated as an ensemble of $N$ free point charges with a charge density $\rho(\vec{r})$. Besides, the influence of the surface charge~$\Phi_{\mathcal{S}}(\vec{r}\,)$ is denoted by a surface charge density $\sigma_{\mathcal{S}}$. The electrostatic approximation $\vec{E}=-\nabla\Phi$ is adopted, which indicates that we can use Poisson's equation to calculate the electric potential in an IMRP-plasma system. Here, index~$n$ refers to the corresponding electrode that a voltage $u_n$ is applied, 
\begin{align}\label{eqn:Poisson}
      -\nabla \!\cdot\! \left(\varepsilon_{\rm r}(\vec{r} ) \nabla \Phi(\vec{r})\right) &= \rho(\vec{r}), \\
	     \Phi(\vec{r}) &= \begin{cases}
	         0, \quad |\vec{r}|\to\infty\\
	         u_n, \quad  \vec{r} \in \mathcal{E}_n \nonumber\\
	    \end{cases}.
\end{align}
A Green's function $G(\vec{r},\vec{r}^{\,\prime})$ is employed as a suitable tool for such a potential problem. In~\cite{Gong}, the proposed model is mathematically derived in detail, and the Green's function is explicitly solved. The electric potential $\Phi(\vec{r},t)$ inside the simulation domain can be specified as
\begin{align} \label{eqn:solver}
	  \Phi(\vec{r},t) =  \frac{1}{\varepsilon_0}\sum_{i=1}^N q_i\, G(\vec{r},\vec{r}_i(t))+\Phi_{\mathcal{S}}(\vec{r}\,) + \sum_{n=1}^{2} u_n(t)  \Psi_{n}(\vec{r}),
\end{align}
where $\Phi_{\mathcal{S}}(\vec{r}\,)$ is the influence of the surface charge, and the characteristic function~$\Psi_{n}(\vec{r}\,)$ represents the vacuum coupling between the electrodes (See Appendix~\ref{App}).

Similar to the particle-in-cell method, the spectral kinetic method consists of two modules: a particle pusher and a field solver. In the field solver, the force acting on the particles can be calculated, then the position and the velocity of the particles are updated in the particle pusher at each simulation step. The dynamics of plasma particles are described in Hamiltonian, where $\vec{p}_k$ is the conjugate momentum of a random free point charge $q_k$ with a mass $m_k$ to the position~$\vec{r}_k$
\begin{align}\label{eqn:hamiltonian}
\!\!\! H\!\left(\vec{r}_1\ldots\vec{p}_N\right) = 
   \sum_{k=1}^N \frac{\vec{p}_k^2}{2 m_k} + \frac{1}{2\varepsilon_0}\sum_{k=1}^N\sum_{i=1}^N q_k q_i\, G(\vec{r}_k,\vec{r}_i) + 
	\sum_{k=1}^N q_k\!\left(\Phi_{\mathcal{S}}(\vec{r}_k)+\sum_{n=1}^{2} u_n(t)  \Psi_{n}(\vec{r}_k)\!\right) \!,
\end{align}
and the canonical equations of motion are
\begin{align}
&\frac{d\vec{r}_k}{dt} = \frac{\partial H}{\partial \vec{p}_k} = \frac{\vec{p}_k}{m_k},	\\
&\frac{d\vec{p}_k}{dt} =-\frac{\partial H}{\partial \vec{r}_k} = 
      -\frac{1}{\varepsilon_0}\sum_{i=1}^N q_k q_i\, \nabla_k G(\vec{r}_k,\vec{r}_i) - 
	 q_k \nabla_k\Phi_{\mathcal{S}}(\vec{r}_k) - q_k\sum_{n=1}^{2} u_n(t)  \nabla_k \Psi_{n}(\vec{r}_k). \nonumber
\end{align}

The input of the simulation is the applied voltage at the electrodes, whereas the output is related to the charge on the electrodes $Q_n$, which can be determined according to Gauss's law, 
\begin{align}
Q_n &=-\int_{\mathcal{E}_n} \varepsilon_0 \varepsilon_{\rm r}(\vec{r} )   \nabla\Phi(\vec{r})\cdot\! {\rm{d}}\vec{f}.
\end{align}

The description of Green's function in the field solver can be expanded in spherical harmonics as an infinite series. However, due to the spherical design and the electrical antisymmetry, only odd modes in the resonance structure are able to propagate. The electric field solely has components in the $r$- and $\theta$-direction. The component in the $\phi$-direction vanishes as the IMRP is azimuthally symmetric. Furthermore, the theoretical and experimental results in ~\cite{JensBrinkmann_Eigen2014,StyrnollAwa2013,StyrnollLapkeAwa2014} 
demonstrate the importance of the first resonance: one particular feature of the MRP is the dominating dipole model. Compared to the first resonance peak, the peaks in the higher modes are barely visible. Consequently, the mathematical model with a modified Green's function describing the IMRP is significantly simplified by considering solely the dipole mode.

Based on the spectral kinetic scheme, the relation between the simulated resonance frequency to the electron plasma frequency is defined numerically. Thus,  the proportionality of resonance and plasma frequency can be obtained from
\begin{align}\label{eqn:ne_wpe}
n_{\rm{e}}=\cfrac{\omega_{\rm{pe}}^2 \varepsilon_0 m_{\rm{e}}}{e^2}.
\end{align}

Another important parameter for process control is the collision frequency. Since it denotes the attenuation of the plasma oscillation, the losses inside the plasma can be described by the collision frequency. Hence, the full width at half maximum $\Delta\omega$ of the resonance curve is directly linked to the effective collision rate $\nu_{\rm{eff}}$. It can be assumed that $\nu_{\rm{eff}}$ consists of two collision rates: the collision rate $\nu_{\rm{col}}$ and the kinetic collision rate $\nu_{\rm{kin}}$
\begin{align}\label{eqn:v_eff}
\nu_{\rm{eff}}=\nu_{\rm{col}}+\nu_{\rm{kin}},
\end{align}
where $\nu_{\rm{col}}$ represents the elastic collisions of electrons with neutrals, and $\nu_{\rm{kin}}$ imitates the influence of the kinetic effects. In general, the elastic collisions between electrons and neutrals are dominant due to the low ionization degree in technical plasmas. However, in a low-pressure regime, the kinetic effects cannot be ignored anymore. Therefore, to describe the electrons deflected by the electric field of the IMRP, the kinetic collision rate $\nu_{\rm{kin}}$ is defined, which is dependent on the electron temperature $T_{\rm{e}}$. With the thermal velocity of the electrons $v_{\rm{th,e}}$ and a length scale factor $L$, we can write  
\begin{align}\label{eqn:formula_1}
\nu_{\rm{kin}}=\cfrac{v_{\rm{th,e}}(T_{\rm{e}})}{L}.
\end{align}
It indicates that $\Delta\omega$ of the simulated resonance curve is of particular interest evaluating $T_{\rm{e}}$. The underlying relation is determined by altering the electron temperature in the simulation.
\newpage
\section{Implementation of the model}
In the simulation of IMRP, the radius is defined as $R = 0.004\,\rm{m}$, and without the dielectric, $R_{\rm{E}} = 0.003\,\rm{m}$ is given. It is convenient to introduce dimensionless notation: $r\to R\,r$, $t\to {\omega_{\rm{pe}}^{-1}}\,t$, $q\to e\,q$, $m\to m_{\rm{e}}\, m$, and $n\to n_{\infty} \,n$. Here, $n_{\infty}$ is defined as the electron density at infinity, where the potential is assumed $0$. Depending on the investigated plasma ($T_{\rm{e}}$, $n_{\rm{e}}$), by solving the Poisson-Boltzmann equation, the stationary state can be calculated as the initial condition of the simulation (See Appendix~\ref{App.Initial}). Therefore, the static density profiles for ions and electrons are numerically determined, in addition to the static surface charge. Notably, the system fulfills the charge equilibrium. Thus, the particles are loaded according to these calculations. In reality, the energy distribution function can be complicated. However, as a general case, all the generated particles follow the Maxwellian velocity distribution function. The setting of simulation is discussed in detail in~\cite{Gong}.

Due to the symmetry of the model, the input and output of the simulation can be simplified. The asymmetry part of the applied voltage ($\Delta u  = u_1-u_2$) contributes to the input of the simulation, while the symmetry part ($\bar{u} = u_1+u_2$) represents the floating potential. A small signal is only provided at the first time step to demonstrate a behavior similar to an impulse to all the particles. Then the linear behavior of the plasma is recorded as the output, which is described by the charge difference ($\Delta Q(t)  = Q_1(t)-Q_2(t)$) on the electrodes. Since the influence of the input signal is solely limited at the beginning of the simulation, the heating phenomena can be neglected. Therefore, the diffuse reflection of the particles is considered. That is, the particle is reflected back with the same speed in a random direction once it reaches the boundary.


In~\cite{Lapke2008,Lapke2011,Lapke2013}, the IMRP-plasma system in the Durde model is converted into an equivalent lumped element circuit, which is simplified since the probe is electrically antisymmetric and geometrically symmetric. Then the distinct dipole resonance peak can be characterized by the real part of the admittance $Y$. In the experiment, the admittance is derived from the measured system's input reflection coefficient $S_{11}$~\cite{M.Fiebrandt}. Therefore, to obtain the resonance curve as the impulse response in the spectral kinetic model, a fitted analytical function is defined to match the numerical output, then the real part of the admittance in the frequency domain can be derived by applying a Fourier transform. The mathematical expression is given as
\begin{align}
	  f(t)=  a e^{-b t} \sin(c t),
\end{align}
where $a$ is the amplitude, $b$ is the damping factor and $c$ is the resonance frequency.

\begin{figure}[!htb]
\includegraphics[width=0.85\textwidth]{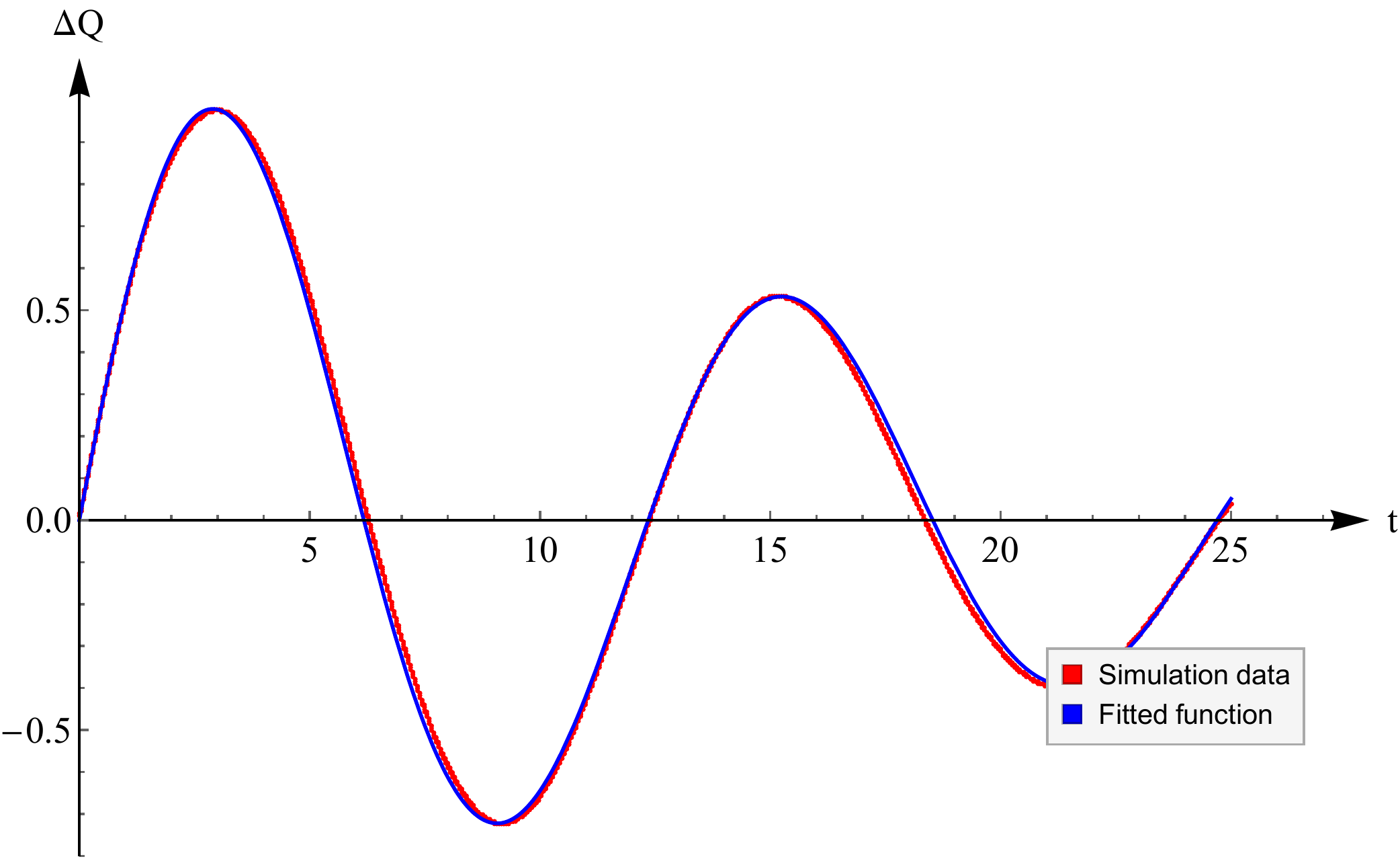}  
	\caption{A fitted function (Blue) is defined to match the simulation data (Red).}
	\label{fig:Fitting}
  \end{figure}

Fig.~\ref{fig:Fitting} shows the example of a plasma with $T_{\rm{e}}=2\,\rm{eV} $ and $n_{\rm{e}}=1.0\times10^{15}\,\rm{m}^{-3}$. The fitted analytical function is determined as
\begin{align}
	  f(t)=  1.135 e^{-0.049 t} \sin(0.509 t),
\end{align}

\begin{figure}[!htb]
\includegraphics[width=0.85\textwidth]{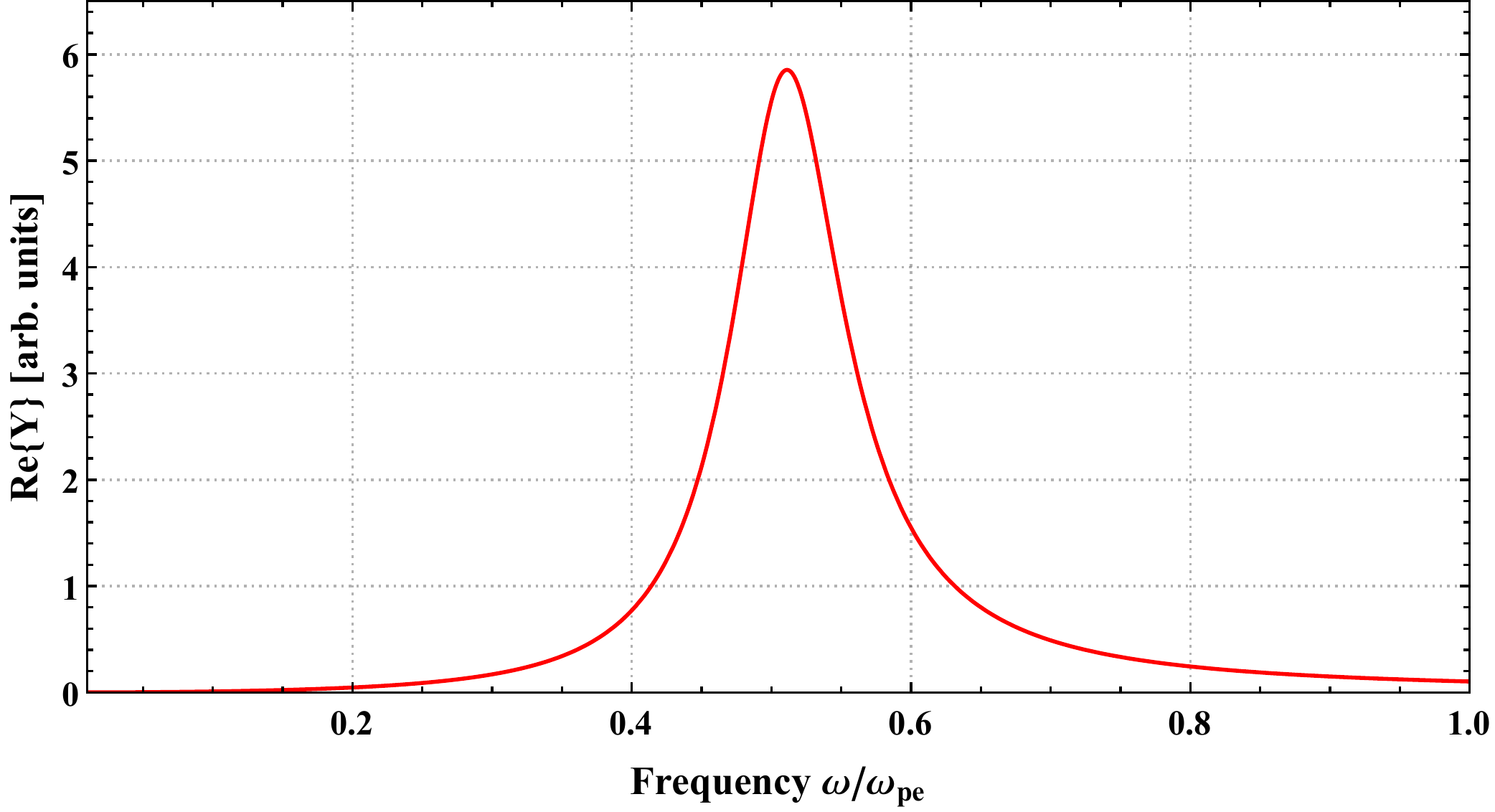}  
	\caption{The corresponding resonance curve is obtained in the frequency domain after the Fourier transformation.}
	\label{fig:FittingY}
  \end{figure}

Applying the Fourier transform, as shown in Fig.~\ref{fig:FittingY}, the resonance curves are determined. Consequently, $\omega_{\rm{res}}$ and $\Delta\omega$ can be obtained from the simulated resonance curves.  

\newpage
\section{Simulation results and discussion}
The spectral kinetic simulation provides the possibility to determine the electron density $n_{\rm{e}}$ (from the resonance frequency $\omega_{\rm{res}}$ of the plasma) and the electron temperature $T_{\rm{e}}$ (from the $\Delta\omega$ of the resonance curve). In this section, simulations in the dipole mode with varying plasma parameters are analyzed. The determination of the electron density and the electron temperature is demonstrated, respectively. Eventually, the comparison between the spectral kinetic model, the Drude model, and the experimental data is presented demonstrating the advantages of the proposed model in the low-pressure plasma.

\subsection{Resonance frequency and electron density}
In \cite{Lapke2008,Lapke2013}, the Drude model is utilized to discuss the relation between the electron density and the resonance frequency. The IMRP-plasma system was described via an infinite number of series resonant circuits. Each resonance mode is represented by a series circuit. The complex admittance $Y$ is evaluated as the plasma dependent frequency response of the system. For the dipole mode, the analytical relation between the electron plasma frequency and the resonance frequency can be derived as
\begin{align}
\omega_{\rm{res}}=\sqrt{\cfrac{2}{3}\left(1-0.82\left(1+\frac{\delta}{R}\right)^{-3}\right)}\omega_{\rm{pe}},
\end{align} 
where $\delta$ represents the sheath thickness. Here, $\delta=3\lambda_{\rm{D}}$ is a reasonable approximation according to~\cite{LiLiu2010}. The direct information of the electron plasma frequency can be provided by adjusting the electron density, then the corresponding resonance frequency is determined.  

\begin{figure}[!htb]
 \includegraphics[width=0.85\textwidth]{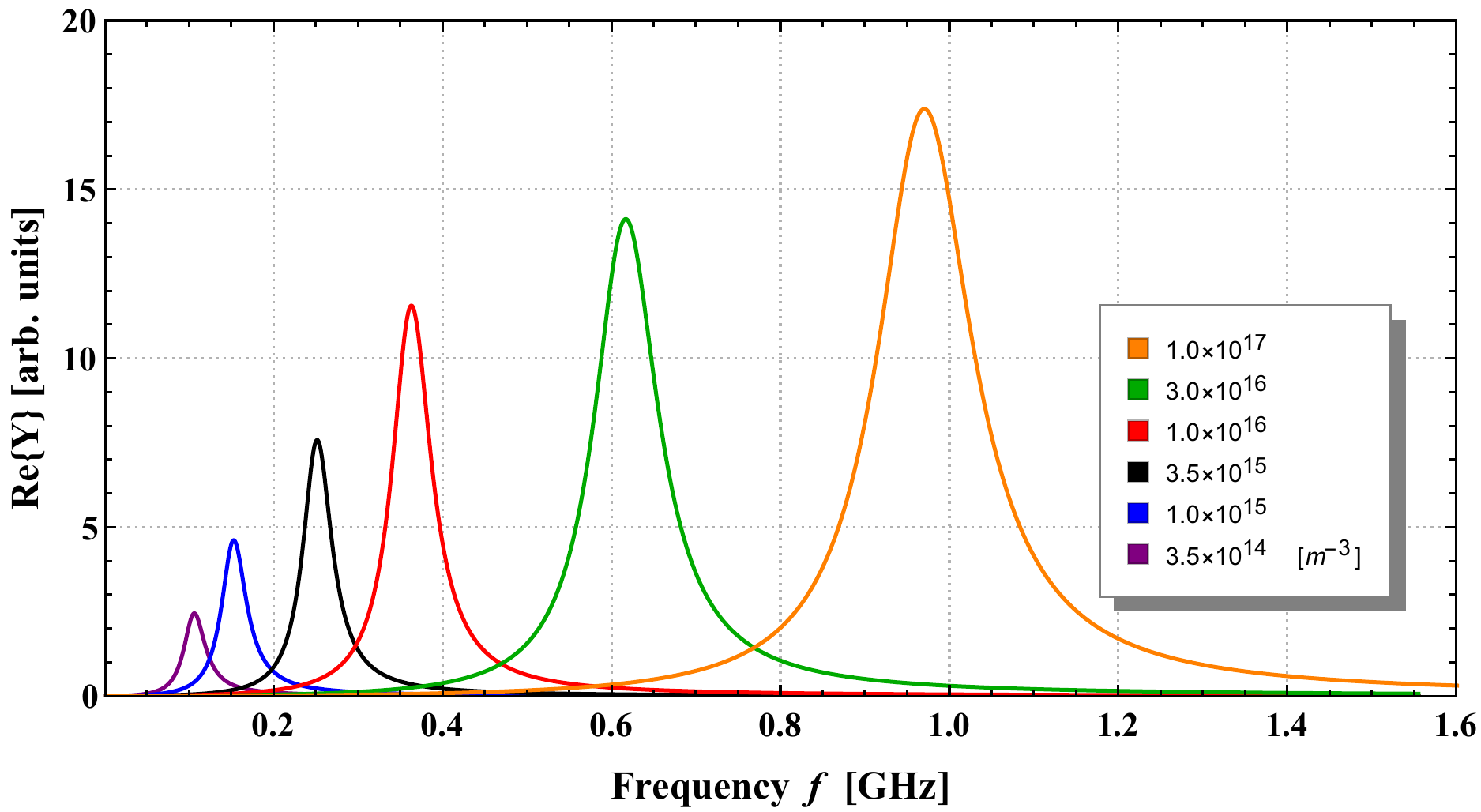}  
	\caption{Simulated resonance behavior of the IMRP for $f_{\rm{pe}}\in\{0.17, 0.28, 0.53, 0.9, 1.56, 2.84\}$ GHz, corresponding to a variation of the plasma density between $3.5\times10^{14}\rm{m}^{-3}$ to $1.0\times10^{17}\rm{m}^{-3}$.}
 \label{fig:Diff_ne}
\end{figure}

\begin{figure}[!htb]
 \includegraphics[width=0.85\textwidth]{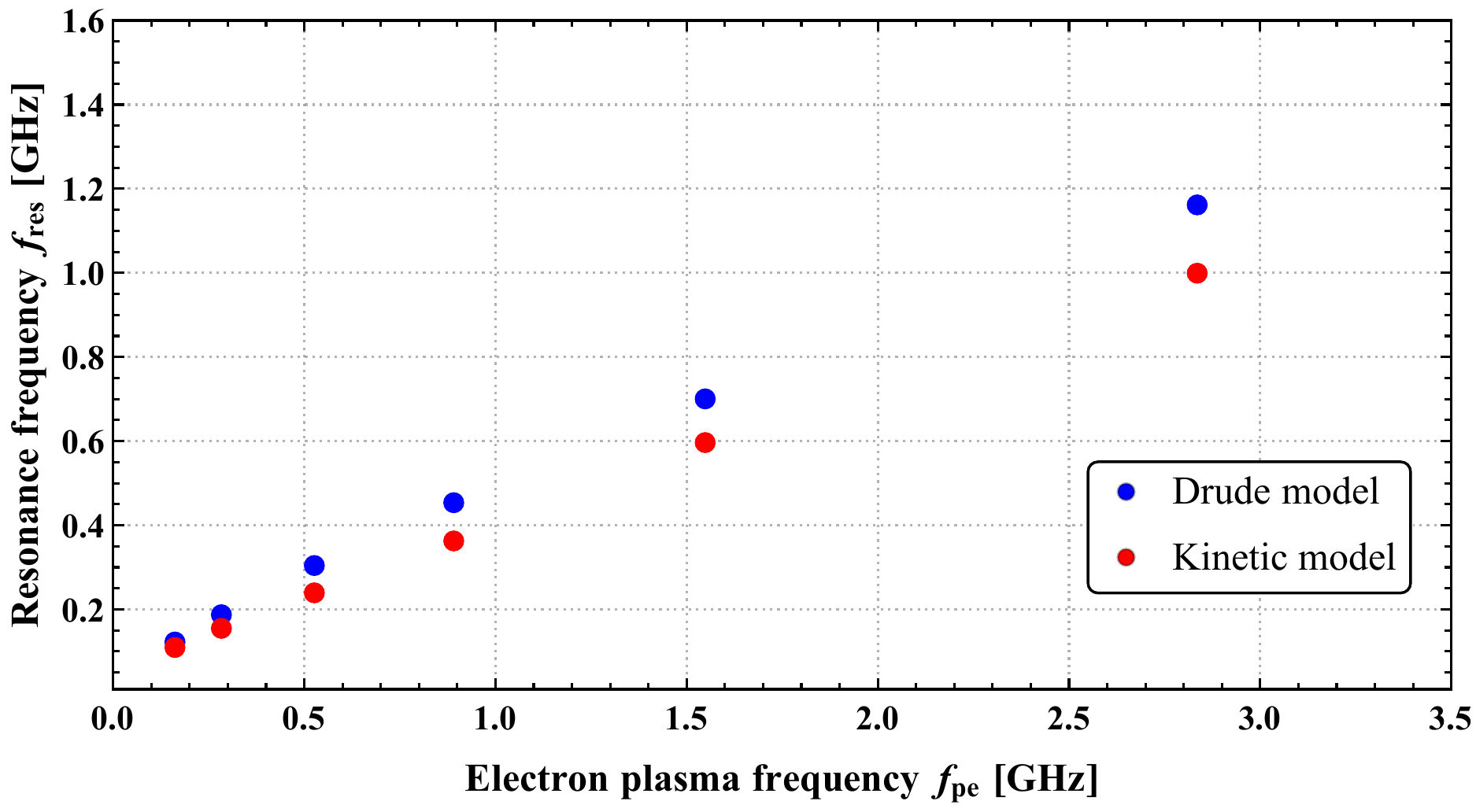}
	\caption{Comparison of the resonance frequencies between the Drude model (blue dot) and the kinetic model (red dot) for a variation of the electron plasma frequencies between 0.17 and 2.84 GHz.}
	\label{fig:fres_fpe}
  \end{figure}

Moreover, the same phenomenon is predicted by the spectral kinetic model. For a rising electron density $n_{\rm{e}}$, a rising resonance frequency $\omega_{\rm{res}}$ is observed in the kinetic simulation. This relation is numerically obtained for a variation of the considered electron density. Since the collision frequency is irrelevant to the resonance frequency, we can set $\nu_{\rm{col}}=0$. With $\omega_{\rm{pe}}=2\pi f_{\rm{pe}}$, the simulation results of the collisionless cases with different electron densities are presented in Fig.~\ref{fig:Diff_ne}. Here, the electron temperature is fixed at $T_{\rm{e}}=3\, \rm{eV}$, and the electron densities are given in $10^{16}\,\rm{m}^{-3}$: $0.035, 0.1, 0.35, 1.0, 3.0, 10.0$. The resulting resonances for the corresponding electron plasma frequencies $\omega_{\rm{pe}}=2\pi\cdot\{0.17, 0.28, 0.53, 0.9, 1.56, 2.84\}\cdot ~10^9\,\rm{s}^{-1}$ are recorded, including the information of the resonance frequency and the normalized \linebreak admittance. The increase of the resonance frequency $f_{\rm{res}}$ is demonstrated for a rising electron density $n_{\rm{e}}$. 
As a result, the calculated resonance frequency in the Drude model (blue dot) and the simulated resonance frequency in the kinetic model (red dot) for the same electron plasma frequency are presented in Fig.~\ref{fig:fres_fpe}. A relatively reasonable agreement can be achieved. Although the results for higher electron plasma frequencies show a divergence in the resonance frequency, the resonance behavior is comparable. In fact, considering that the assumption of a sheath thickness is not necessary for this self-consistent particle-based model, the spectral kinetic scheme can be seen as a reliable approach to predict the resonance frequency based on the electron density.

 \subsection{$\Delta\omega$ and electron temperature}
The FWHM $\Delta\omega$ of the resonance curve describes the damping of the IMRP-plasma system, which can be determined to evaluate the effective collision frequency $\nu_{\rm{eff}}$.  In the low-pressure regime, the kinetic effects dominate the total energy loss. The particles are deflected by the field of the IMRP, and the energy is transported out of the perturbed domain of the plasma. In order to reveal the influence of the kinetic effects on $\Delta\omega$, the simulation of the collisionless case with different electron temperatures is investigated. The assumption of collisionlessness ($\nu_{\rm{col}}=0$) indicates that $\nu_{\rm{eff}}$ is identical to $\nu_{\rm{kin}}$, where $\nu_{\rm{kin}}$ is proportional to $\Delta\omega$. The thermal velocity can be written as $v_{\rm{th,e}}=  \lambda_{\rm{D}}(T_{\rm{e}})\,\omega_{\rm{pe}}$, where the corresponding Debye length $\lambda_{\rm{D}}$ depends on the electron temperature. Assuming a length scale $L=R$, the formula~(\ref{eqn:formula_1}) can be expressed with an unknown coefficient $k$ as
\begin{align}
\Delta\omega=k\cdot\cfrac{\lambda_{\rm{D}}(T_{\rm{e}})\omega_{\rm{pe}}}{R}.
\end{align} 

\begin{figure}[!htb]
 \includegraphics[width=0.85\textwidth]{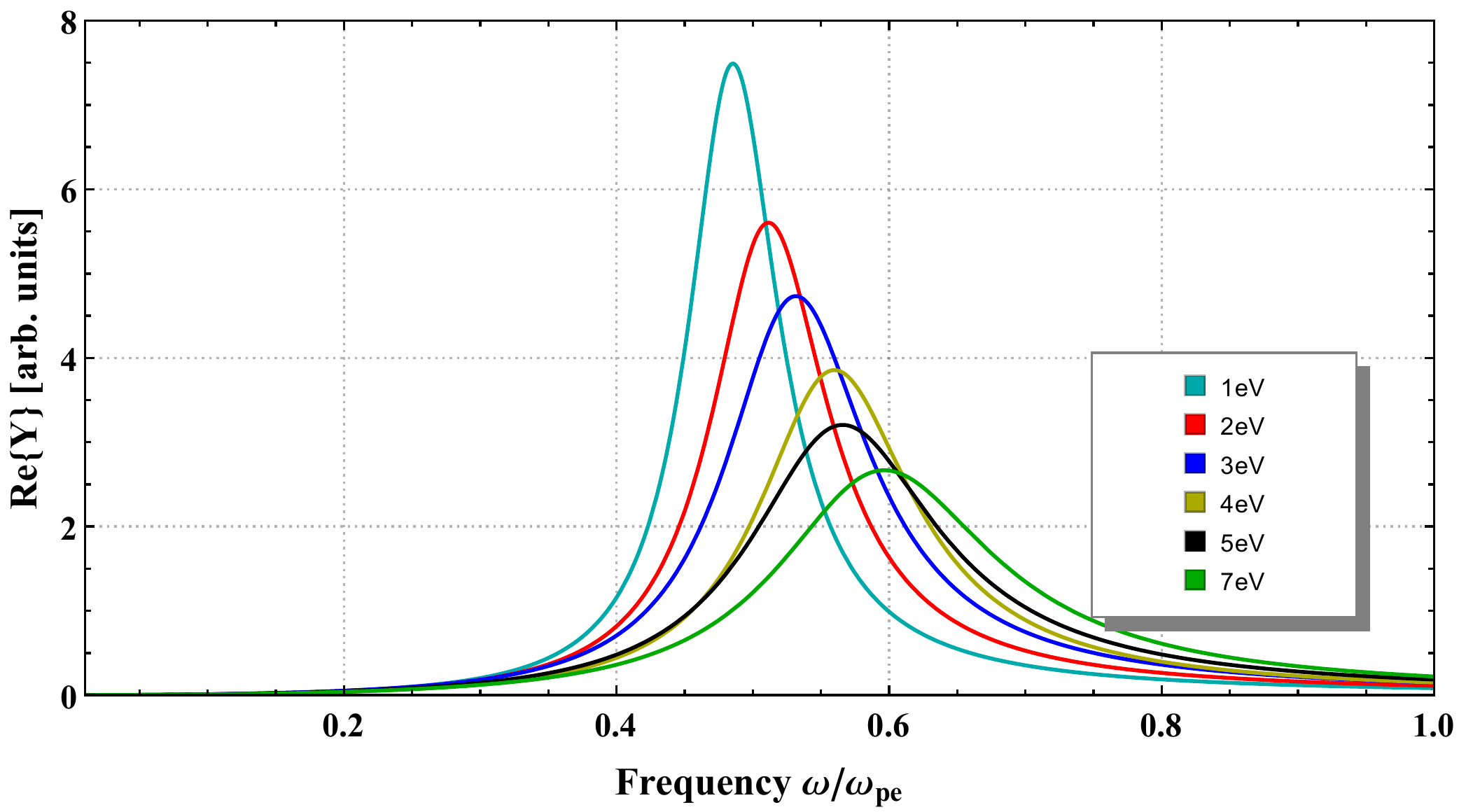}  
  \caption{Simulation results of the collisonless spectral kinetic simulation of the IMRP for a variation of the electron temperature $T_{\rm{e}}$ at a constant electron density $n_{\rm{e}}=1\times10^{15}\,\rm{m}^{-3}$.}
 \label{fig:diff_Te}
\end{figure}

\begin{figure}[!htb]
 \includegraphics[width=0.85\textwidth]{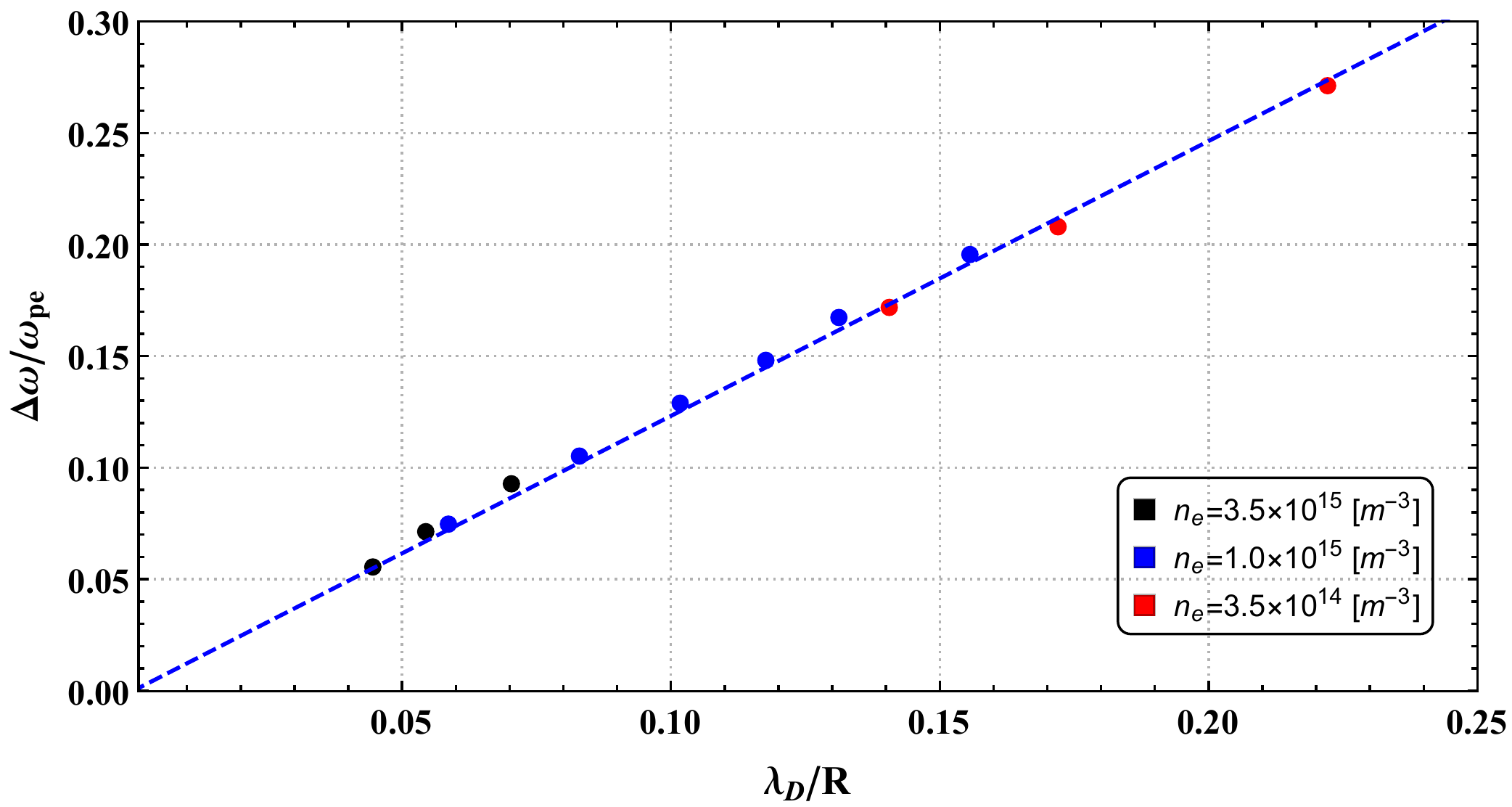}  
  \caption{Evaluated FWHM $\Delta\omega$ of the simulated resonance curves with different plasma parameters.}
 \label{fig:DeOm_Lam}
\end{figure}

Taking the plasma with an electron density $n_{\rm{e}}=1\times10^{15}\,\rm{m}^{-3}$ as an example. In Fig.~\ref{fig:diff_Te}, the simulated resonance curves for a variation of the electron temperature $T_{\rm{e}}\in \{1, 2, 3, 4, 5, 7\}\, \rm{eV}$ are presented. It shows that a rising electron temperature leads to the broadening of the resonance curve due to pure collisionless kinetic effects. Then the FWHM are determined as $\Delta\omega\in\{0.075, 0.105, 0.129, 0.148, 0.167, 0.194\}\,\omega_{\rm{pe}}$. As shown in Fig.~\ref{fig:DeOm_Lam}, the calculated $\Delta\omega/\omega_{\rm{pe}}$ (blue dot) can be approximated as a linear function of $\lambda_{\rm{D}}/R$, where the coefficient $k$ is the slope. Thus, this coefficient is calculated as $k=1.264$ at this specific investigated electron density. Similarly, such a formula can be derived from the simulation results for a variation of the electron density. For the plasma with $n_{\rm{e}}=3.5\times10^{14}\,\rm{m}^{-3}$ and $n_{\rm{e}}=3.5\times10^{15}\,\rm{m}^{-3}$, simulations of $T_{\rm{e}}=2\,\rm{eV}$, $3\,\rm{eV}$, and $5\,\rm{eV}$ are executed, respectively. The results (black and red dot) agree well with the proposed formula. The coefficient yields at $k=1.232$. Eventually, within a certain magnitude the proposed formula can be obtained as  
\begin{align}
\Delta\omega(T_{\rm{e}})=\cfrac{k}{R}\sqrt{\frac{T_{\rm{e}}}{m_{\rm{e}}}}.
\end{align}

As stated, such a linear relation was speculated in~\cite{Lapke2008,SchulzStyrnollLapke2012} for over a decade, which is verified in the parameter study of the spectral kinetic model. Strictly speaking, there is small deviation due to the weakly dependence of the electron density to the slope $k$. Therefore, a modified formula is expected to cover such second-order effects in further investigation. However, with the prediction of $\Delta\omega$ from the investigated plasma using spectral kinetic simulation, in practice, a lookup table can be created to determine the desired electron temperature from the measured FWHM.

 \subsection{Comparison of $\Delta\omega$} 
\begin{figure}[!htb]
\includegraphics[width=0.85\textwidth]{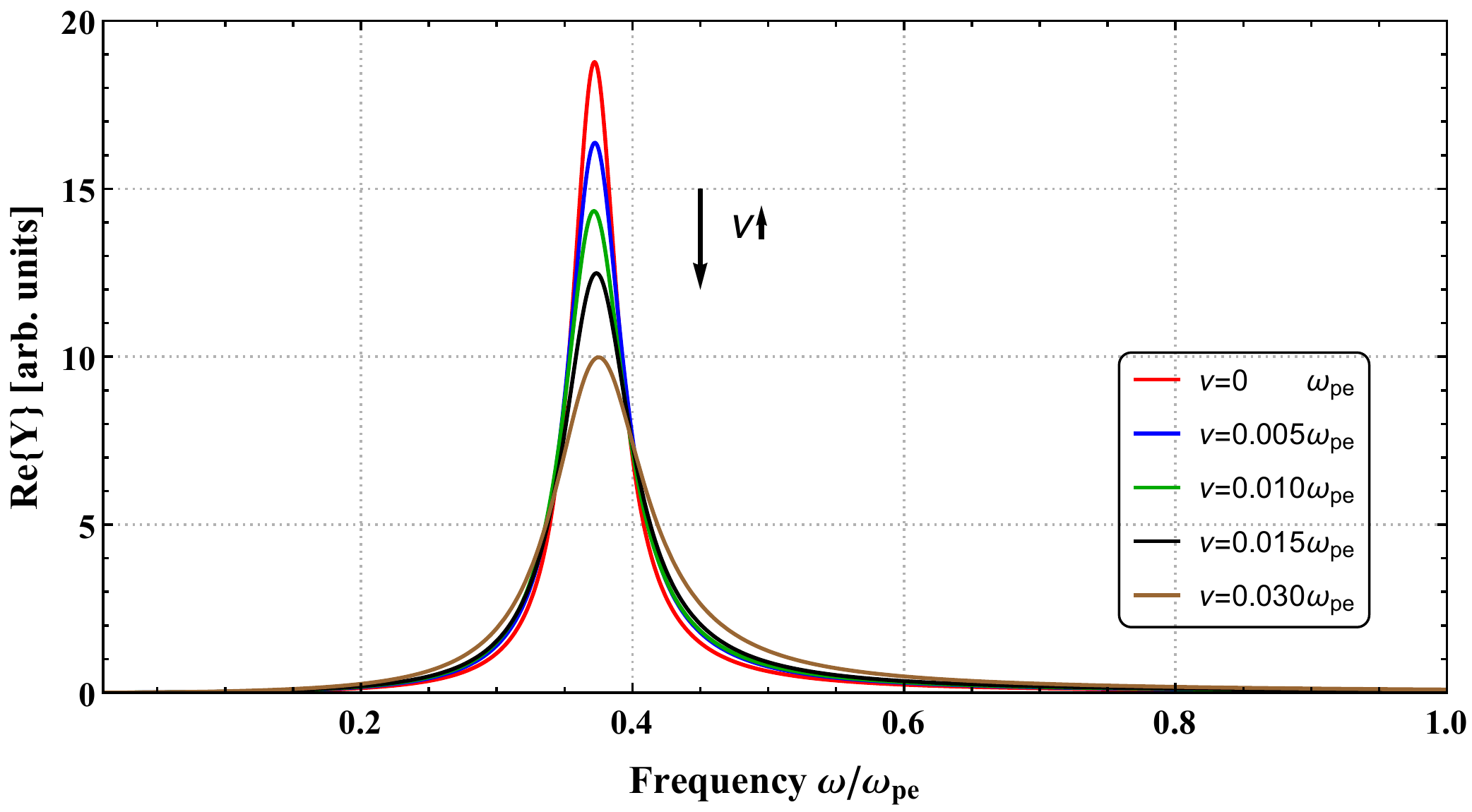}  
	\caption{Simulated resonance curve with varying collision frequencies for the plasma with $T_{\rm{e}}=3\,\rm{eV}$ and $n_{\rm{e}}=3\times10^{16}\,\rm{m}^{-3}$.}
	\label{fig:Incr_Col_Freq}
  \end{figure}

For a complete investigation of the IMRP-plasma system, besides kinetic effects, the elastic collisions between electrons and neutral atoms need to be considered in the total energy loss. The relation between the collision frequency and resonance width is discussed in~\cite{SchulzCollision2016}, where the plasma is considered via the Drude model and the absence of kinetic effects is covered by a compensated electron frequency. Here, the kinetic effects are well demonstrated via the spectral kinetic simulation. Fig.~\ref{fig:Incr_Col_Freq} shows the comparison of the simulated resonance curve for varying electron-neutral collision frequencies $\nu\in\{0, 0.005, 0.01, 0.015, 0.03\}\,\omega_{\rm{pe}}$. In this case, the plasma is with electron density $n_{\rm{e}}=3\times10^{16}\,\rm{m}^{-3}$ and electron temperature $T_{\rm{e}}=3\,\rm{eV}$. The increase of electron-neutral collision frequency results in a broadening of the simulated resonance curve. It is introduced that $\Delta\omega$ represents the damping in the system, which\linebreak consists of the influence of the kinetic effects ($\nu_{\rm{kin}}$) and the energy loss caused by the elastic collisions ($\nu_{\rm{col}}$). As expected, the kinetic effects play an important role in the low-pressure regime. FWHM is numerically determined as $\Delta\omega$ $\in\{0.424, 0.492, 0.549, 0.61, 0.81\}\,\omega_{\rm{pe}}$. Therefore, the total energy loss can be feasibly predicted using the spectral kinetic model. 
  
The measurement of the MRP-plasma system with the same investigated plasma parameters is given in~\cite{Jens_thesis}. That is, the constant electron density is at $n_{\rm{e}}=3\times10^{16}\,\rm{m}^{-3}$, and the electron temperature is $T_{\rm{e}}=3\,\rm{eV}$. The setup is a cylindrical double inductively coupled plasma reactor (diameter is $40 \,\rm{cm}$, and height is $20 \,\rm{cm}$). The plasma parameters are validated by reference measurements with a Langmuir probe~\cite{Godyak2011}. Then, the plasma process of MRP is controlled via excitation power and gas pressure. A network analyzer is employed to measure the complex reflection coefficient. Afterward, the real part of the admittance is calculated from the obtained input reflection. The resonance frequency is held constant by fixing the excitation power, and the comparable collision frequency is achieved by the adjustment of the gas pressure~\cite{Measurement,M.Fiebrandt}. 
 

To compare the simulation results and the measurements, converting the collision frequency to the pressure is required. As discussed earlier, collisions are the fundamental process in the plasma. The general discussion is given by~\cite{Lieberman2005}
\begin{align}
\nu_{\rm{col}}= \frac{p_{\rm{gas}}}{k_{\rm{B}}\,T_{\rm{gas}}}\cdot K(T_{\rm{e}}),
\end{align}
where $T_{\rm{gas}}$ is the temperature of the neutral background, $p_{\rm{gas}}$ represents the gas pressure, and $k_{\rm{B}}$ is the Boltzmann's constant. $K(T_{\rm{e}})$ states that the elastic collision frequency $\nu_{\rm{col}}$ is a function of $T_{\rm{e}}$. According to~\cite{Gud_Coll}, the elastic collision frequency in argon for $T_{\rm{e}}$ between $1\rm{eV}$ and $7\rm{eV}$ is approximated by
\begin{align} 
\nu_{\rm{col}}= \cfrac{p_{\rm{gas}}}{k_{\rm{B}}T_{\rm{gas}}} \exp\left[{-31.388+1.609 \ln\left(\frac{T_{\rm{e}}}{{\rm{eV}}}\right)+0.062\ln\left(\frac{T_{\rm{e}}}{{\rm{eV}}}\right)^2-0.117\ln\left(\frac{T_{\rm{e}}}{{\rm{eV}}}\right)^3}\right].
\end{align}

\begin{figure}[!htb]
\includegraphics[width=0.85\textwidth]{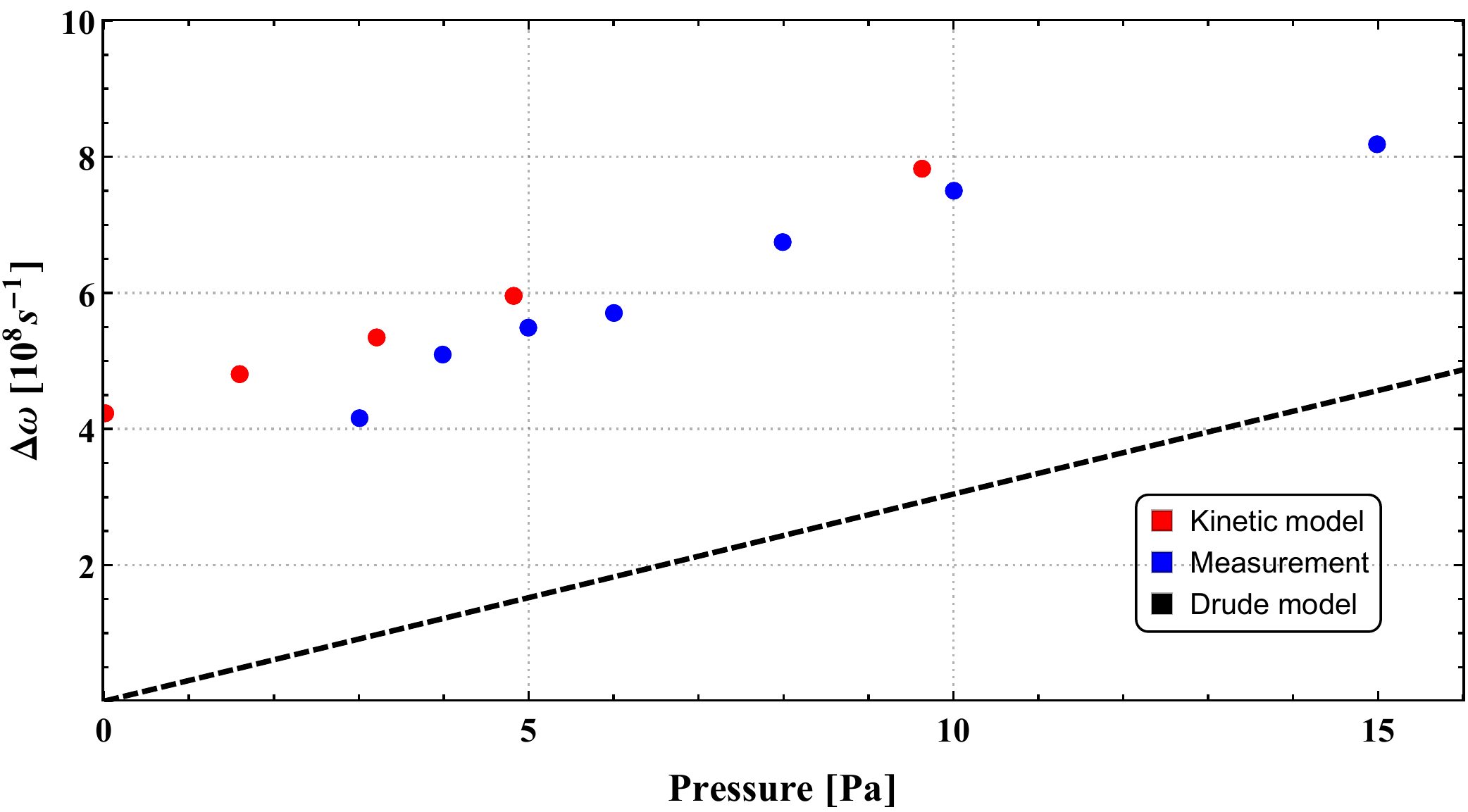}  
	\caption{Comparison of the FWHM $\Delta \omega$ between the measured spectra, the simulated results in the spectral kinetic model, and the electron-neutral collision frequency in the Drude model.}
	\label{fig:Sim_Exp_Comp}
  \end{figure}

Consequently, the comparison of $\Delta\omega$ for different pressure is presented in Fig.~\ref{fig:Sim_Exp_Comp}. It shows that $\Delta\omega$ of the measured spectra (blue dot) is comparable to the simulated results of the spectral kinetic model (red dot). The results based on the Drude model (black dashed line) can be derived directly from the given collision frequency due to the absence of the kinetic collision frequency. Notably, it is difficult to measure $\Delta\omega$ in the low pressure. The kinetic results indicate the possible errors in the measurements, especially in the lower pressure. However, it is demonstrated in the comparison that the offset caused by the limitation of the Drude model is covered in the spectral kinetic simulation. Furthermore, a good agreement between the kinetic simulation and the measurement confirms the suitability and reliability of the proposed spectral kinetic model.


\newpage
\section{Summary and conclusion}
In this paper, we introduce a plasma probe based on active plasma resonance spectroscopy, the MRP, and the spectral kinetic simulation of the IMRP is discussed. The real part of the admittance of the dipole mode is determined for a variation of the electron density and the electron temperature. As expected, kinetic effects, such as the broadening of the resonance curve and damping phenomena, are observed in the simulation results. It describes the energy loss due to the escape of the free particles from the perturbed domain within the plasma, which is absent in the Drude model. The influence of those effects is particularly emphasized in the low-pressure regime. Moreover, the possibility to simultaneously evaluate the electron density and the electron temperature from the resonance curve is explored in a parameter study of the proposed model. 

A relation between the resonance frequency and the electron plasma frequency is obtained in the kinetic simulation, which is comparable to results via the Drude model. The desired electron density can be derived from the corresponding electron plasma frequency. Besides, the determination of $\Delta\omega$ from the simulated resonance curve is discussed, which is directly linked to the energy loss. The formula to describe the relation between $\Delta\omega$ and $T_{\rm{e}}$ is proposed in the previous study. However, it has not been explicitly clarified. In this work, the aforementioned formula is verified. Based on the spectral kinetic model, the simulation of the collisionless cases determines the pure kinetic collision frequency, which is of particular importance in evaluating the electron temperature. In other words, the kinetic collision frequency can be expressed as a function of $T_{\rm{e}}$. Considering the fact that the elastic electron-neutral collisions are also dependent on $T_{\rm{e}}$, we are able to determine the desired electron temperature from the measured resonance curve. 

Eventually, the spectral kinetic simulation shows a good agreement with the measurement. The captured energy losses are comparable in the pressure $p\,\leq10\,\rm{Pa}$, which indicates the reliability and suitability of the proposed model. Besides, compared to the Drude model, the more realistic physics can be described by the kinetic model. The explanation for the offset in the Drude model is found, which is due to the absence of kinetic effects. Consequently, the spectral kinetic model is applicable to predict the plasma parameters.

 \newpage
\section{Citation}

\newpage
\begin{appendix}

\section{The formal solution of Poisson's equation}
\label{App}   
The Poisson's problem is to be solved in IMRP-plasma system. The electric potential can be calculated in Poisson's equation~\eqref{eqn:Poisson}. A suitable tool for the formal description is the Green's function $G(\vec{r},\vec{r}^{\,\prime})$, which is defined as 
\begin{alignat}{2}\label{eqn:Green function}
    - \nabla  \cdot\! \left( \varepsilon_{\rm r}(\vec{r} ) \nabla  G(\vec{r},\vec{r}^{\,\prime})\right)
	&= \delta^{(3)}(\vec{r}\,-\vec{r}^{\,\prime}),\\
	   G(\vec{r},\vec{r}^{\,\prime})& = 0, \quad   \vec{r} \in \mathcal{E} \quad \text{or} \quad |\vec{r}|\to\infty. \nonumber
\end{alignat}

By applying the Green's second identity, the formal solution $\Phi(\vec{r})$ of the Poisson equation can be established. As is suggested by the boundary conditions, the Green's function vanishes at the electrodes and infinity. Then the potential can be written as
\begin{align}\label{eqn:App potential}
	  \Phi(\vec{r},t) =   \frac{1}{\varepsilon_0} \sum_{i=1}^N q_i\, G(\vec{r},\vec{r}_i(t))+ \frac{1}{\varepsilon_0}\int_{\mathcal{S}} \sigma_{\mathcal{S}} \, G(\vec{r},\vec{r}^{\,\prime})\, {\rm{d}}f^{\,\prime} \nonumber \\
 +  \int_{\mathcal{E}} \Phi(\vec{r}^{\,\prime}) \varepsilon_{\rm r}(\vec{r}^{\,\prime})  \nabla^\prime G(\vec{r},\vec{r}^{\,\prime})  \cdot\! {\rm{d}}\vec{f}^{\,\prime}, 
\end{align}
which is contributed by the free particles, the surface charge, and the influence of the electrodes. Here $\mathcal{E}$ represents all electrodes.

The latter formula is defined as the characteristic function $\Psi_{n}(\vec{r}\,)$, where index $n=1,2$ depends on the electrodes. It contains the information about the geometry, which is independent of the plasma
\begin{align} 
	  \Psi_{n}(\vec{r}\,) = \int_{\mathcal{E}}  \varepsilon_{\rm r}(\vec{r}^{\,\prime})  \nabla^\prime G(\vec{r},\vec{r}^{\,\prime})  \cdot\! {\rm{d}}\vec{f}^{\,\prime}. 
\end{align}

Replacing $\Phi_{\mathcal{S}}(\vec{r})$ to the surface term in~\eqref{eqn:App potential}, the potential can be simplified as 
\begin{align} 
	  \Phi(\vec{r},t) =  \frac{1}{\varepsilon_0}\sum_{i=1}^N q_i\, G(\vec{r},\vec{r}_i(t))+\Phi_{\mathcal{S}}(\vec{r}\,) + \sum_{n=1}^{2} u_n(t)  \Psi_{n}(\vec{r}).
\end{align}

\newpage
\section{Initial condition of the simulation}
\label{App.Initial}
The initial condition of the simulation is related to a spherically symmetric probe-plasma equilibrium under floating conditions. The Poisson equation can be expressed with the electron and ion densities
\begin{align}\label{eqn:app_Poisson}
      - \varepsilon_0\frac{1}{r^2}\frac{\partial}{\partial r} r^2  \frac{\partial\Phi}{\partial r} = e(n_{\rm{i}}(r)-n_{\rm{e}}(r)).
\end{align}
The electron density can be obtained by the Boltzmann relation. Assuming the potential at infinity is 0 where the plasma is quasi-neutral with density $n_{\infty}$.
\begin{align}\label{eqn:app_density_ele}
      n_{\rm{e}}=n_{\infty} \exp\left(\frac{e\Phi}{T_{\rm{e}}}\right).
\end{align}

The ion flux to the probe is spatially constant, which can be written as a product of $n_{\infty}$, the Bohm velocity $\sqrt{T_{\rm{e}}/m_{\rm{i}}}$, and an unknown constant $\hat{R}$.
\begin{align}
     4\pi r^2n_{\rm{i}}v_{\rm{i}}= -  4\pi\hat{R}^2 n_{\infty} \sqrt{\frac{T_{\rm{e}}}{m_{\rm{i}}}}.
\end{align}

With the energy conservation, the ion velocity and density can be obtained
\begin{align}
      \frac{1}{2}m_{\rm{i}} v_{\rm{i}}^2 + e \Phi = 0.
\end{align}

The ion density is then calculated as
\begin{align}\label{eqn:app_density_ion}
      n_{\rm{i}}=\frac{\hat{R}^2}{r^2}
      \,n_{\infty}\sqrt{-\frac{T_{\rm{e}}}{2e\Phi}}.
\end{align}

Inserting \ref{eqn:app_density_ele} and \ref{eqn:app_density_ion} into \ref{eqn:app_Poisson}, we have
\begin{align}\label{eqn:app_diff_eqn}
      - \varepsilon_0\frac{1}{r^2}\frac{\partial}{\partial r} r^2     \frac{\partial\Phi}{\partial r}  =  e n_{\infty} \left(\cfrac{\hat{R}^2}{r^2}\sqrt{-\frac{T_{\rm{e}}}{2e\Phi}}-\exp\left(\frac{e\Phi}{T_{\rm{e}}}\right) \right).
\end{align}

The floating condition indicates that the ion flux equals to the electron flux on the probe. 
\begin{align}
    4\pi\hat{R}^2 n_{\infty} \sqrt{\frac{T_{\rm{e}}}{m_{\rm{i}}}}
    = 4\pi R^2
      \sqrt{\frac{T_{\rm{e}}}{2\pi m_{\rm{e}}}}\, n_{\infty} \exp\left(\frac{e\Phi(R)}{T_{\rm{e}}}\right).
\end{align}
By using shooting method, the unknown constant $\hat{R}$ can be determined depending on the investigated plasma parameters. Then the potential and density profiles of the electrons and ions are obtained as the initial condition of the simulation.

\end{appendix}

\newpage
\section*{Figures}

\newpage
 
\newpage

\newpage

\newpage
 
\end{document}